\documentclass[a4paper]{article}
\usepackage{Odyssey2022}
\usepackage{epsfig,amssymb,amsmath}
\ninept
\usepackage{multirow}
\usepackage{listings}
\usepackage{amsmath}
\usepackage{url}
\usepackage{xcolor}
\usepackage{booktabs}

\definecolor{codegreen}{rgb}{0,0.6,0}
\definecolor{codegray}{rgb}{0.5,0.5,0.5}
\definecolor{codepurple}{rgb}{0.58,0,0.82}
\definecolor{backcolour}{rgb}{0.95,0.95,0.92}

\lstdefinestyle{mystyle}{
    backgroundcolor=\color{white},   
    commentstyle=\color{codegreen},
    keywordstyle=\color{magenta},
    numberstyle=\tiny\color{codegray},
    stringstyle=\color{codepurple},
    basicstyle=\ttfamily\small,
    breakatwhitespace=false,         
    breaklines=false,                 
    captionpos=b,                    
    keepspaces=true,                 
    numbers=left,                    
    numbersep=5pt,                  
    showspaces=false,                
    showstringspaces=false,
    showtabs=false,                  
    tabsize=2
}

\title{Collar-aware Training for Streaming Speaker Change Detection in Broadcast Speech}
\name{Joonas Kalda, Tanel Alumäe}
\address{Department of Software Science  \\
Tallinn University of Technology, Estonia \\
{\small \tt joonas.kalda@taltech.ee, tanel.alumae@taltech.ee} }

\begin{document}
%\ninept
%
\maketitle
\begin{abstract}
In this paper, we present a novel training method for speaker change detection models. Speaker change detection is often viewed as a binary sequence labelling problem. The main challenges with this approach are the vagueness of annotated change points caused by the silences between speaker turns and imbalanced data due to the majority of frames not including a speaker change. Conventional training methods tackle these by artificially increasing the proportion of positive labels in the training data. Instead, the proposed method uses an objective function which encourages the model to predict a single positive label within a specified collar. This is done by marginalizing over all possible subsequences that have exactly one positive label within the collar. Experiments on English and Estonian datasets show large improvements over the conventional training method. Additionally, the model outputs have peaks concentrated to a single frame, removing the need for post-processing to find the exact predicted change point which is particularly useful for streaming applications.
\end{abstract}
%
%\begin{keywords}
%Speaker change detection, sequence labelling
%\end{keywords}
%
\section{Introduction}
\label{sec:intro}

Speaker change detection (SCD) is a task of locating precise points in the audio recording when a different speaker starts speaking. It is often used as the first step in speaker diarization systems. Depending on the application, SCD systems can be either streaming (also known as \textit{online}) or batch-processing (\textit{offline}). In a batch processing system, the whole audio recording is available when SCD is applied. This allows the model to use all information from both past and future frames when locating speaker change points. A streaming model, on the other hand, needs to identify speaker change points with low latency, using typically only one or two seconds of audio from the future. Streaming SCD is needed as a preprocessing step in streaming speech recognition systems that perform unsupervised speaker adaptation, e.g. using i-vectors \cite{saon2013speaker}, so that the speaker adaptation state could be reset at speaker change points. SCD is also often an explicit requirement in realtime closed captioning systems for broadcast television \cite{aronowitz2020context}.

Most modern SCD systems are based on supervised learning. Large speech datasets, manually annotated with speaker change points, are used for training and SCD is treated as a binary sequence classification task. Long short-term memory (LSTM) recurrent neural networks \cite{india2017lstm, yin2017speaker} or convolutional neural networks \cite{hruz2016convolutional, hruz2017convolutional, mateju2019approach} are often used as models. An important issue when training such models for SCD is that the annotated change points in the training data are ambiguous and imbalanced. The ambiguousness comes from the fact that often there is a substantial amount of silence between the speech of two adjacent speakers, yet only a single frame is marked as a change point. The choice where exactly the annotated change point resides is often inconsistent, resulting in training data that is confusing for the model. Also, the number of frames in the training data labelled as change points is usually less than 1\% of all the frames, causing problems with model convergence.

In this paper, we propose a novel objective function for training sequence classification models for SCD. This \textit{collar-aware} objective function gives the SCD model more freedom by allowing it to choose an appropriate speaker change point within the neighbourhood of the annotated change point. This method addresses both the problems of imbalanced data as well as the ambiguousness of the annotated labels. Furthermore, the models trained using this method are especially well suited for streaming applications, as the resulting model generates ``peaky'' change points that do not require any post-processing to find local maxima. We show that the method also achieves notably higher accuracy in both streaming and batch-processing scenarios, compared to several well-established baselines\footnote{Code and demo available at \url{https://github.com/alumae/online_speaker_change_detector}}.

\section{Related work}

SCD approaches can be divided into two main categories: metric- and model based. The first approach operates by applying a pair of sliding windows on the sequence of feature vectors  extracted from the underlying audio signal and uses a divergence metric for comparing their contents. A speaker change point is detected if the divergence between two adjacent windows is larger than a predefined threshold and the divergence achieves a significant local maximum. 
The advantage of this method is that it doesn't require a large annotated training corpus for training: only the value of the threshold parameter needs finetuning on a small validation set. This method is used in many speaker diarization systems that  use Gaussian mixture models (GMMs) as their main building blocks (e.g. \cite{rouvier2013open})

A model based approach, on the other hand, uses a a training corpus with manually annotated speaker change points to train a model for this task. Many different models have been proposed, such as hidden Markov models \cite{meignier2001hmm}, GMMs \cite{malegaonkar2007efficient}, eigenvoices \cite{castaldo2008stream}, deep neural networks (DNNs) \cite{gupta2015speaker, mateju2019approach}, convolutional neural networks \cite{hruz2016convolutional, hruz2017convolutional, mateju2019approach}, recurrent neural networks \cite{india2017lstm, yin2017speaker} and Siamese networks \cite{sari2019pre}.

Since models based on neural networks have become popular in recent years for this task, we review three approaches based on them more carefully. In \cite{yin2017speaker}, SCD is formulated as a standard binary sequence labelling task that can be tackled using bidirectional LSTMs: the model's task is to label each frame with either 0 (no speaker change) or 1 (speaker change). One problem with this approach is that the training data is heavily imbalanced: the number of frames that are labelled with 0 is  much larger than the number of frames labelled with 1 (only 0.4\% according to \cite{yin2017speaker}). Under standard training, the model converges to a state in which a 0 is predicted for each frame. Therefore, \cite{yin2017speaker} increases the number of positive labels artificially by labelling frames 50 ms on each side of the annotated change point as 1. During inference, local score maxima exceeding a pre-determined threshold are marked as speaker change points. 

In \cite{mateju2019approach}, a somewhat similar approach is used, but instead of a bidirectional LSTM, a CNN is used that ``sees'' a fixed-size window of feature frames prior and after the current frame. This allows operating the model with low latency in streaming mode. As with the LSTM-based approach, a large number of frames in the direct neighbourhood of the annotated change point are labelled as positive during model training, in order to make the training data more balanced. 

In \cite{sari2019pre}, a Siamese architecture is used for low-latency SCD: a 2-second window prior and after the current frame is processed by a bidirectional LSTM, resulting in two embedding vectors. The embeddings are then processed by a classification module that decides whether the two segments correspond to different speakers. Various pretraining schemes can be applied to the embedding computation module that are found to improve the detection performance by a large amount. 
This work handles the imbalanced data problem by sampling a predefined ratio of speaker change points from the training data to each batch.

Inconsistent and unreliable speaker turn boundaries in manually annotated training data can also have a negative effect on the performance of end-to-end speaker diarization systems. In \cite{zeghidour2021dive}, a modification to the standard multilabel classification loss for speaker diarization is introduced that simply ignores the errors in a defined radius around annotated speaker change points.

\section{Collar-aware training}

\begin{figure}[tb]
  \centering
  \includegraphics[width=0.99\linewidth]{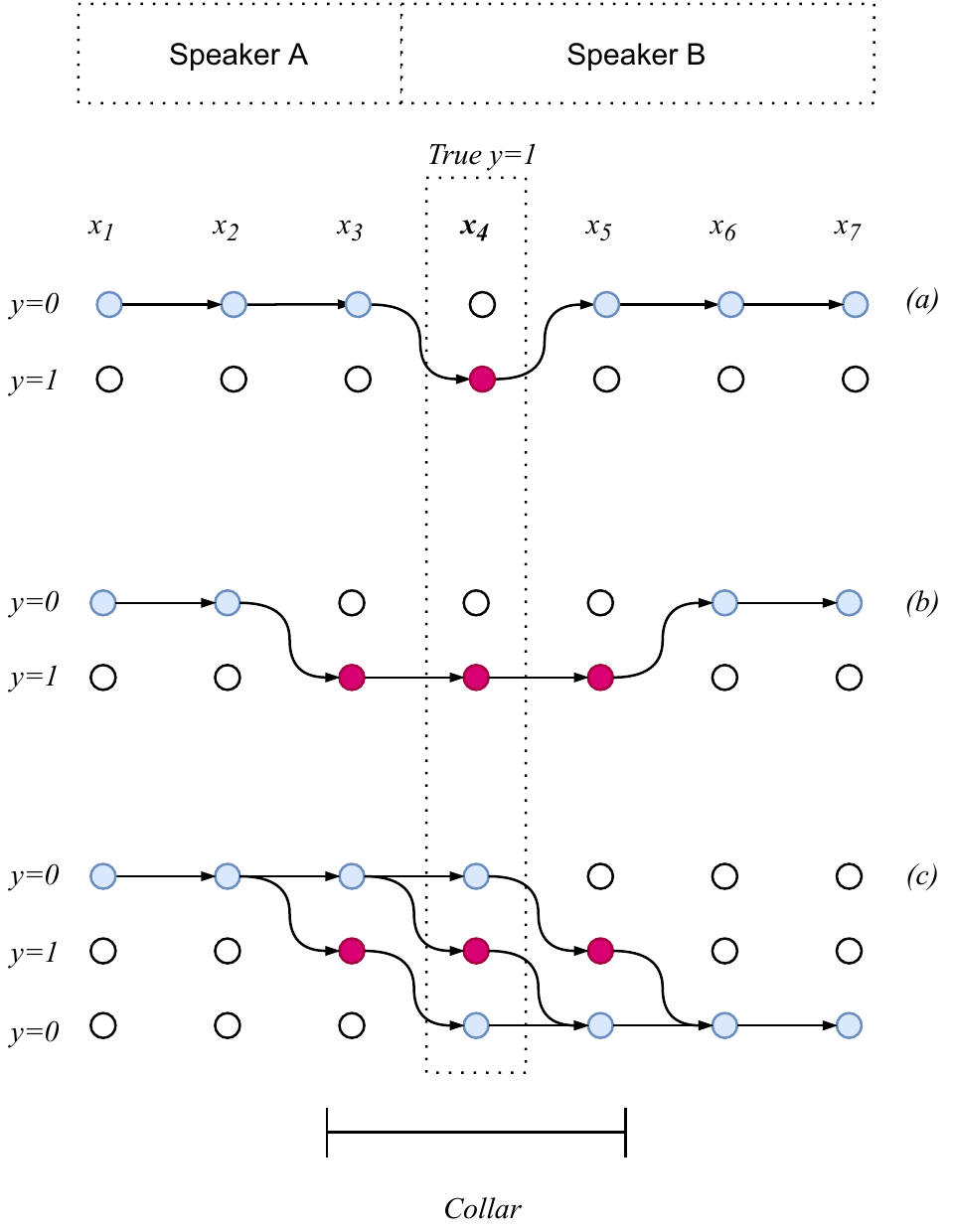}
  \caption{\textbf{Outline of three supervision methods for speaker change detection:} \textit{(a)} corresponds to 
  standard sequence labelling objective; \textit{(b)} increases the number of positive
  labels artificially by setting several frames in the neighbourhood of the annotated change point as positive \cite{yin2017speaker}; \textit{(c)} the 
  proposed method sums over all paths that have exactly one positive label in the neighbourhood of the annotated change point.}
  \label{fig:training}
\end{figure}

\begin{figure*}
\begin{lstlisting}[language=Python]
def collar_bce_loss(log_probs, change_points, collar):
  """
  Compute collar-aware binary CE loss.
  
  Arguments:
  log_probs -- tensor of shape (seq_len, 2), containing log likelihoods of non-boundary 
    and boundary events 
  change_points -- indexes of annotated boundaries
  collar -- value of the collar (in frames)
  """
  result = log_probs[:, 0].sum()
  for change_point in change_points:
    collar_variant_logs = []
    collar_start_i = change_point - collar
    collar_end_i = change_point + collar
    time_index = range(collar_start_i, collar_end_i + 1)
    event_index = torch.eye(collar_end_i - collar_start_i + 1).long()
    collar_variant_logprobs = log_probs[time_index, event_index].sum(1)
    result -= log_probs[time_index, 0].sum()
    result += torch.logsumexp(collar_variant_logprobs, 0)        
  return -result
\end{lstlisting}
  \caption{Pytorch code for efficient calculation of the collar-aware binary cross-entropy loss.}
  \label{fig:code}
\end{figure*}

Speaker change detection is often regarded as a binary sequence labelling problem.
We consider an audio recording consisting of feature vectors $x_i$ for $i = 1, ..., N$ and the corresponding speaker boundary labels $y_i \in  \{0,1\}$ with $y_i = 1$ meaning that the frame corresponds to an annotated speaker boundary. 

When a SCD system is evaluated in terms of precision and recall of detected speaker boundaries, it is a standard practice to use a \textit{collar}  (typically 250 ms) for annotated speaker boundaries: if the boundary detected by the model is within the tolerated amount of milliseconds of the annotated boundary, the detected speaker change point is assumed to be correct. However, under standard sequence labelling objective (Figure \ref{fig:training}, \textit{a}), the collar is not used, making the training objective different from the evaluation scenario. 

As pointed out in the previous section, several papers \cite{yin2017speaker, mateju2019approach} have suggested to artificially modify the training data of the speaker boundary detection model by labelling a predefined number of frames around the annotated speaker boundary as additional (pseudo-)boundaries (Figure \ref{fig:training}, \textit{b}). This is done in order to make the training data more balanced in terms of label frequencies, and to 
model the inherent ambiguousness of the speaker boundaries. 

We propose to use a modified objective function for training SCD models that solves both the problems of imbalanced data and ambiguous annotated boundaries. Instead of labelling points around the annotated boundary as pseudo-boundaries, it supervises the model to label exactly one frame within the given collar as a speaker boundary, but the exact position of the boundary can be freely chosen  (Figure \ref{fig:training}, \textit{c)}. This method has several advantages: (1) it matches the evaluation criteria better than method (\textit{b}); (2) it solves the imbalanced data problem similarly or better than method (\textit{b}); (3) the model trained in this manner can be easily applied in online mode without any post-processing to find the local maximum, since the output of the model is now very ``peaky'' (see Section \ref{peakiness}).

\begin{figure}[tb]
  \centering
  \includegraphics[width=0.99\linewidth]{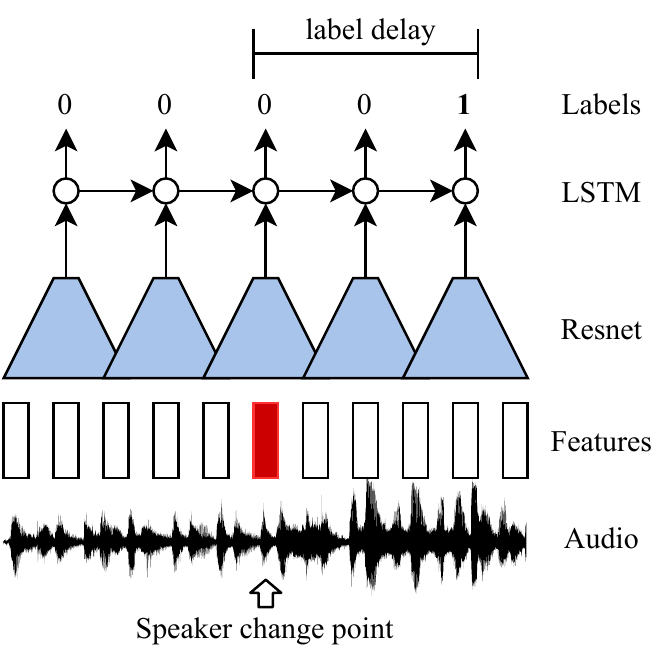}
  \caption{\textbf{Architecture of the streaming \textit{Resnet-LSTM} model:} Filterbank features from speech frames are fed into a Resnet module that is pretrained on a speaker recognition task and then trained jointly with the rest of the model. The outputs from the Resnet module are fed into a LSTM layer that identifies speaker boundaries ($\mathtt{label}=1$). Since this cannot be done without the knowledge of future frames, the identification is done with a label delay that corresponds to 1 second of speech (in streaming mode).}
  \label{fig:architecture}
\end{figure}

Formally, given the reference labels $y$ and model predictions $\hat{y_i}$, the standard binary sequence labelling objective is:
\[
\mathcal{L}(\hat{y}, y) = - \sum_{i=1}^{N}y_i \log(\hat{y_i}) + (1-y_i)\log(1- \hat{y_i})
\]

Since the boundaries occur very sparsely, this objective can be efficiently calculated by summing over the log likelihoods of the no-boundary events, and then modifying it to account for the few boundary events.
Given annotated boundary positions $Z = \{z | y_z=1\}$, the standard sequence labelling loss becomes:
\begin{equation*}
\begin{split}
\mathcal{L}(\hat{y}, Z) = - \Big( & \sum_{i=1}^{N}\log(1- \hat{y}_i) \\
  -  & \sum_{z_i \in Z} \log(1- \hat{y}_{z_{i}}) + 
\sum_{z_i \in Z} \log(\hat{y}_{z_{i}}) \Big)
\end{split}
\end{equation*}
In order to calculate the proposed collar-aware objective, we have to consider a superset $S(Z)$ of all sets of boundary events $Z'$ where for each original change point $z_i \in Z$ there is exactly one boundary event that is within it's collar set $C_i = \{x | z_i - c < x < z_i + c\}$, where $c$ is the value of the collar. Alternatively,
\[
S(Z) = \{\{z_1,...,z_N\} | z_i \in C_i \forall i \in \{1,...,N\}\},
\]
\sloppy
where $N$ is the total number of original change points. For example, if the reference label sequence  is $[0 0 1 0 0]$, then $Z'$ at ${\rm collar}=2$ corresponds to a set of label sequences $\{[0 1 0 0 0], [0 0 1 0 0], [0 0 0 1 0]\}$.

%In order to calculate the proposed collar-aware objective, we have to consider a superset $S(Z)$ of all sets of boundary events $Z'$ where for each change point in the original $Z$ there is exactly one boundary event that is within the collar of the original change point: $Z' = \{z'_i | z_i - c < z'_i  < z_i + c\}$, where $c$ is the value of the collar. For example, if the reference label sequence  is $[0 0 1 0 0]$, then $Z'$ at ${{\rm collar}=1$ corresponds to a set of label sequences $\{[0 1 0 0 0], [0 0 1 0 0], [0 0 0 1 0]\}$.
The proposed objective sums over all such change point configurations:
\[
\mathcal{L}_{\mathrm{collar}}(\hat{y}, Z) = - \log \sum_{Z' \in S(Z)} \mathrm{e}^{-\mathcal{L}(\hat{y}, Z')}
\]
% \[
% - \log \sum_{z' \in \{2,3,4\}} e^{-\mathcal{L}(\hat{y}, z')}
% \]
% \[
% \mathcal{L}_{collar}(\hat{y}, Z) = - \sum_{Z' \in S(Z)} \log \sum_{z' \in Z'} e^{-\mathcal{L}(\hat{y}, \{z'\})}
% \]
The idea of this objective function is somewhat similar to the CTC loss function \cite{hannun2017sequence} used for training end-to-end speech recognition models. As with CTC, it is not practical to compute it using brute force. In order to make it more efficient, we can again use our knowledge that the number of speaker boundaries occur very sparsely. Figure \ref{fig:code} lists the Pytorch implementation of this idea.  The collar-aware loss can be calculated by summing over the log-likelihoods of the non-boundary events (line 11), subtracting the log likelihoods of the non-boundary events that lie within a boundary collar (line 19), and then adding the marginalized log-likelihood of having exactly one boundary somewhere within the collar (line 20).

\fussy

\section{Experiments}

\subsection{Datasets}

The experiments were carried on both English and Estonian datasets. For English, we used the HUB4 speech dataset \cite{hub4_a, hub4_b}. The Estonian dataset consists of TV and radio broadcasts. Both datasets are manually transcribed and annotated with speaker information.
Test and development data were separated similarly for both datasets: 10 recordings were chosen for each at random. An overview of dataset sizes and annotated boundary counts is provided in Table \ref{tab:data}. The datasets are similarly balanced, with 0.04\% of the frames being labelled as speaker change points.

\begin{table}[]
\caption{A comparison of lengths and the number of speaker change points in the datasets used in the experiments.}
\begin{tabular}{l|ccc}
\toprule
Dataset         & Train         & Development       & Test      \\ 
\midrule
Estonian & 497.2h / 80k   & 1.2h  / 166 & 0.7h  / 102 \\
English  & 128.3h / 19.5k & 6.1h  / 893 & 5.4h  / 893 \\
\bottomrule
\end{tabular}
\label{tab:data}
\end{table}

\subsection{Implementation details}

%\subsubsection{Architectures}
We consider two different architectures which we train using both the standard training method and the proposed collar-aware one.

The first architecture was chosen to closely resemble that of \cite{yin2017speaker}. 33-dimensional acoustic features are extracted every 10ms on a 25ms window, consisting of 11-dimensional MFCCs and their first and second derivatives. The model is made up of two Bi-LSTM layers having 64 outputs and 40 outputs and a multi-layer-perceptron with 40-, 10- and 1-dimensional layers. 

The second architecture uses a Resnet-based feature extractor before a LSTM layer. The Resnet module is extracted from a speaker recognition model pretrained on VoxCeleb2 \cite{Chung18b}, as described in \cite{alumae2020taltech}. It results in 1280-dimensional features with a frame subsampling rate of 8. In the low-latency streaming model, the Resnet layer is followed by two 256-dimensional LSTM layers, and  1-second label delay is used in order for the model to see the data past the current frame (see Figure \ref{fig:architecture}). In the offline model, the LSTMs are replaced with bidirectional LSTMs, and no label delay is used.

%\subsubsection{Training methods}

All our training methods use extracted segments with random lengths between 10s and 30s.

The first training method used also follows \cite{yin2017speaker}. Namely the training data is artificially modified by positively labelling every frame in a 50ms neighborhood of an annotated change point. Notably, no additional labelling is needed for the Resnet-based architecture since the subsampling that happens during feature extraction results in frames of 80ms duration and thus a 50ms neighborhood corresponds to roughly a single frame. A standard binary sequence labelling objective is used as the loss function for this method.

The second training method includes no artificial labelling and instead uses the proposed collar-aware objective as the loss function. The size of the collar was chosen to be $c=250$ms and the effects of varying the collar size are discussed in Section \ref{tuning}. 

During training, data augmentation is applied: background noise and/or reverberation is added to each training segment, both with a probability of 0.3. The background noises originate from the MUSAN corpus \cite{musan2015}. For reverberation, we used simulated small and medium room impulse responses \cite{ko2017study} and real room impulse responses from the  BUT Speech@FIT Reverb Database \cite{szoke2019}.

\begin{table*}[tb]
\caption{Precision (P), recall (R) and F1 results of various batch-mode and streaming models on Estonian and English datasets with two two different forgiveness collar values.}
\label{tab:main}
\begin{tabular}{l|ccc|ccc||ccc|ccc}
\toprule
                                                                          & \multicolumn{6}{c||}{Estonian dataset}                                & \multicolumn{6}{c}{English dataset}                                 \\
\midrule
                                                                          & \multicolumn{3}{c|}{collar=0.25s} & \multicolumn{3}{c||}{collar=0.50s} & \multicolumn{3}{c|}{collar=0.25s} & \multicolumn{3}{c}{collar=0.50s} \\
\midrule
Model                                                                     & P    & R   & F1     & P    & R   & F1     & P    & R   & F1     & P    & R   & F1     \\
\midrule
\textit{Batch-mode processing}        \\
\midrule
Pretrained speaker diarization (VBx)                                      & 0.68         & 0.68     & 0.68   & 0.96         & 0.96     & \textbf{0.96}   & 0.48         & 0.64     & 0.55   & 0.67         & 0.88     & 0.76   \\
Pretrained \textit{pyannote.audio}                   & 0.62         & 0.73     & 0.67   & 0.68         & 0.79     & 0.73   & 0.42         & 0.38     & 0.40   & 0.57         & 0.51     & 0.54   \\
+ finetuned on the given dataset                                          & 0.82         & 0.82     & 0.82   & 0.89         & 0.89     & 0.89   & 0.60         & 0.49     & 0.54   & 0.73         & 0.59     & 0.65   \\
BLSTM                                                                      & 0.7  & 0.85 & 0.77 & 0.74 & 0.88 & 0.81  & 0.44         & 0.59     & 0.50   & 0.50         & 0.62     & 0.55   \\
+ collar aware training                                                   & 0.75 & 0.81 & 0.78 & 0.86 & 0.80  & 0.83      & 0.59         & 0.57     & 0.58   & 0.61         & 0.61     & 0.61   \\
Resnet + BLSTM                                                            & 0.80         & 0.78     & 0.79   & 0.84         & 0.80     & 0.82   & 0.59         & 0.66     & 0.62   & 0.65         & 0.69     & 0.67   \\
+ collar aware training                                                   & 0.92         & 0.89     & \textbf{0.91}   & 0.96         & 0.92     & 0.94   & 0.76         & 0.69     & \textbf{0.73}   & 0.79         & 0.76     & \textbf{0.78}   \\
\midrule
\textit{Streaming processing }        \\
\midrule
\textit{pyannote.audio} with latency=1.0s & 0.34         & 0.67     & 0.45   & 0.37         & 0.73     & 0.49   &     0.21 & 0.33 & 0.26 & 0.28 & 0.44 & 0.34   \\
+ finetuned on our data                                                   & 0.42         & 0.68     & 0.51   & 0.46         & 0.75     & 0.57   &  0.26 & 0.45 & 0.32 & 0.30  & 0.52 & 0.38       \\
Resnet + LSTM                                                             & 0.73         & 0.73     & 0.73   & 0.76         & 0.75     & 0.76   & 0.56         & 0.62     & 0.59   & 0.58         & 0.71     & 0.64   \\
+ collar aware training                                                   & 0.89         & 0.83     & \textbf{0.86}   & 0.92         & 0.86     & \textbf{0.89}   & 0.66         & 0.71     & \textbf{0.68}   & 0.72         & 0.75     & \textbf{0.74}  \\
\bottomrule
\end{tabular}
\end{table*}

\subsection{Evaluation metrics}

The evaluation metrics are standard precision and recall calculated on the test sets. Predicted change points are considered correct if they match an annotated change point within a forgiveness collar (closest pairs are matched first until no pairs remain).  Although our main evaluation metrics are precision and recall of detected speaker change points at a forgiveness collar of 250 ms, we also show the same metrics using a larger 0.5 second collar. A change point is predicted to happen if the local maximum of the models output is higher than a threshold, the value of which is determined by maximizing the F1 score on the development set.

%mention peak detection

\begin{figure}[tb]
  \centering
  \includegraphics[width=0.99\linewidth]{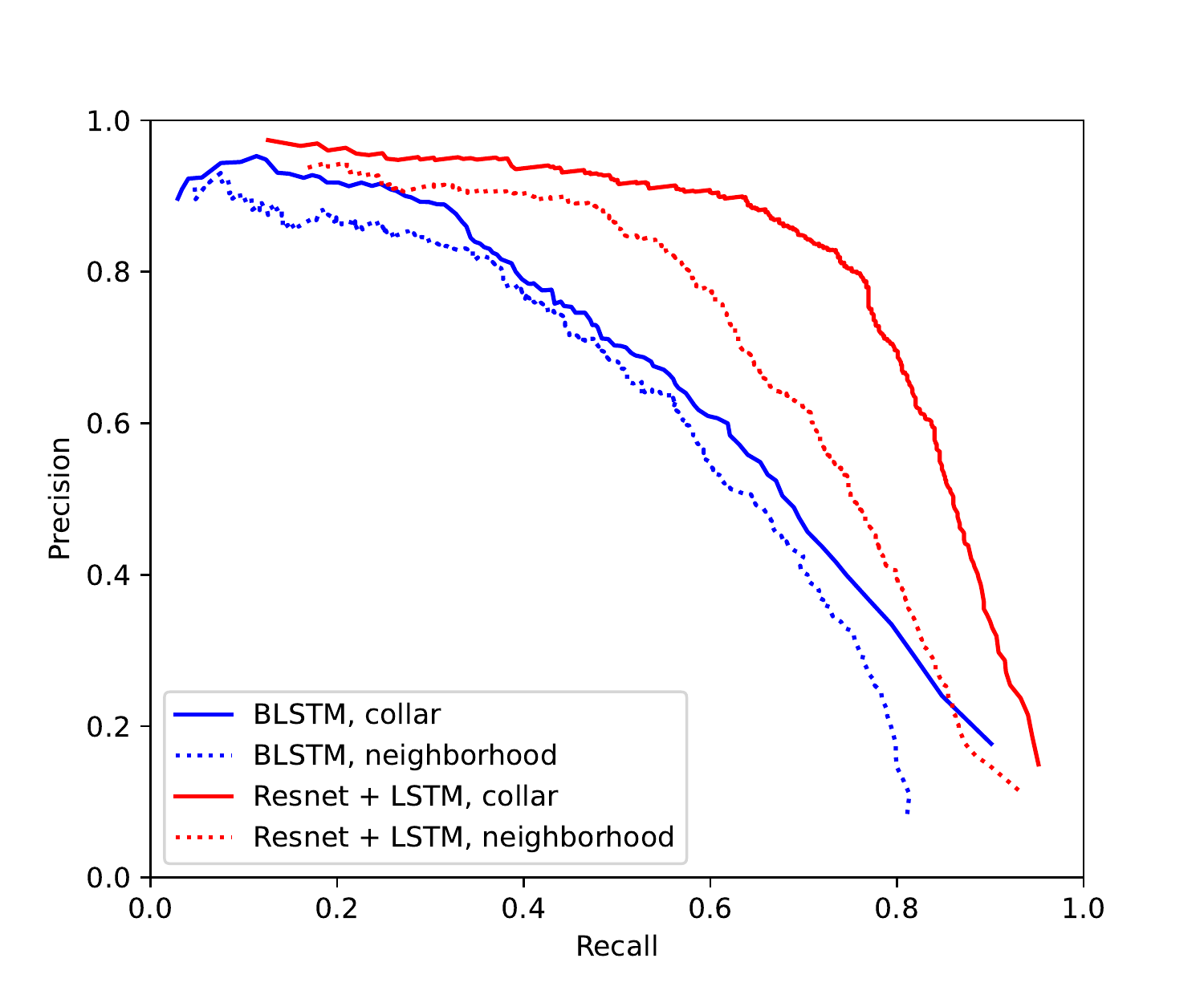}
  \caption{Precision-recall curves for models trained on the English dataset.}
  \label{fig:pr}
\end{figure}

\begin{figure*}[tb]
  \centering
  \includegraphics[width=\linewidth]{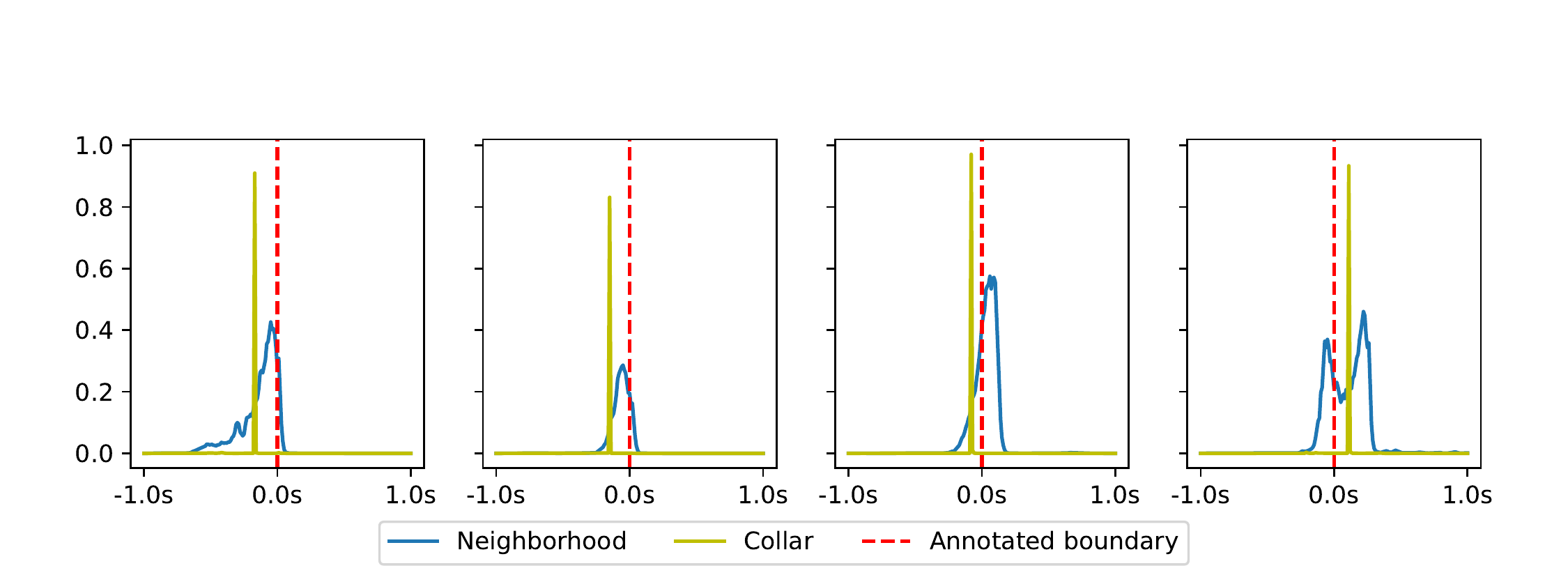}
  \caption{Random samples of Resnet+BLSTM model outputs for neighborhood- and collar-based models trained on the English dataset centered around annotated speaker change points.}
  \label{fig:peakiness}
\end{figure*}

\subsection{Baselines}

In addition to the pure LSTM-based speaker segmentation system \cite{yin2017speaker}, we compare our results to various baselines. Since speaker change points can be easily derived from the output of a speaker diarization system, we use several speaker diarization models that have achieved competitive results on various diarization benchmarks.

The recently proposed VBx diarization method \cite{landini2022bayesian} has produced state-of-the-art results on CALLHOME, AMI and DIHARD II datasets. The method uses a Bayesian hidden Markov model to find speaker clusters in a sequence of x-vectors. We used the open source implementation of the method available at GitHub\footnote{\url{https://github.com/BUTSpeechFIT/VBx}}. The diarization pipeline first extracts x-vectors from the sections of the audio that contain speech. The provided x-vector models are trained on VoxCeleb1 \cite{nagrani2017voxceleb}, VoxCeleb2 \cite{Chung18b} and CN-CELEB \cite{fan2020cn}. The x-vectors are extracted every 0.25 seconds from overlapping sub-segments of 1.5 seconds. The x-vectors are centered, whitened and length
normalized \cite{garcia2011analysis}.  The x-vectors are pre-clustered using agglomerative hierarchical clustering to obtain the initial speaker labels and finally  further clustered using the VBx model. The used the VBx parameters $F_A$, $F_b$ and $P_{loop}$ tuned on the respective development sets in order to minimize the boundary detection F1 score with a 250 ms forgiveness collar.

We also compare to the neural speaker segmentation method implemented in \textit{pyannote.audio} \cite{bredin2021} that performs joint voice activity detection, speaker segmentation and overlapped speech detection.  Similarly to the original EEND approach \cite{fujita2019end}, here speaker segmentation is modeled as a multi-label classification problem using permutation-invariant training. The  model operates on short audio chunks (5 seconds) at a  temporal resolution of every 16 ms and outputs speaker activation probabilities that are stitched together across frames. More specifically, we use the model available at \url{https://huggingface.co/pyannote/segmentation} that is trained on the DIHARD3 corpus \cite{ryant2020third,ryant2020third-b}.
We also experiment with the same model in a low-latency setting \cite{coria2021overlap}, using the open-source implementation\footnote{\url{https://github.com/juanmc2005/StreamingSpeakerDiarization/}}. In streaming mode, the latency of the segmentation output is configurable. To make the results comparable to our streaming model, we used a latency of 1 second.

In addition to using the publicly available \textit{pyannote.audio} segmentation model, we also experimented with finetuning it on our training data. This was done on each dataset and resulted in further baselines for both streaming and offline settings.

In order to convert the output of the diarization systems to speaker change points we consider all the consecutive pairs of speaker segments where the speaker ids differ. If there is less than 2 seconds between the two segments then a speaker change point is predicted at the beginning of the second segment. This was found to give better results than other choices in the gap like the midpoint or the end of the first segment.

\subsection{Results}

A comparison of the model performances on the two datasets is provided in Table \ref{tab:main}. The results are divided into two categories: models that perform change point detection in batch mode, and streaming models. The Resnet+LSTM based models trained using the proposed collar-aware loss function clearly outperform the same models trained using the standard training method on both datasets. Furthermore, these models also provide higher speaker change point detection accuracy than the baseline speaker diarization models. The state-of-the-art VBx diarization model actually results in impressive accuracy at a collar of 0.5 seconds but much lower accuracy at the standard 0.25 second collar. This might be due to the fact that the VBx model uses a relatively large temporal resolution of 0.25 seconds which causes the detected change point to be considered an error if it is off by just one timestep.

Precision-recall curves obtained by varying the classification threshold on the English test set are presented in Figure \ref{fig:pr}. It can be seen that collar-aware training outperforms neighbourhood-based training at all operation points for both LSTM and Resnet-BLSTM based models.

\begin{figure}[tbh]
  \centering
  \includegraphics[width=\linewidth]{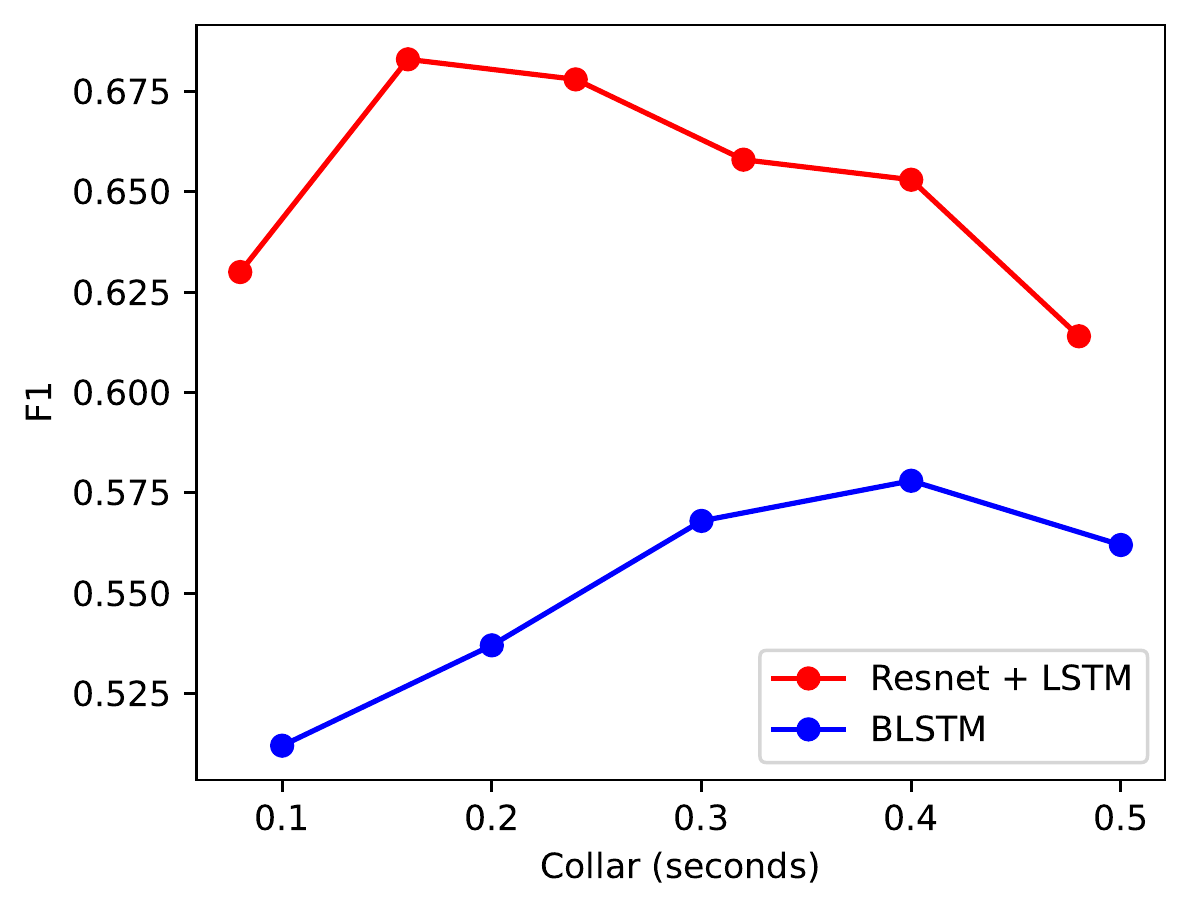}
  \caption{F1 scores on the English dataset for models of varying training collar size.}
  \label{fig:tuning}
\end{figure}

\subsubsection{Peakiness of model output}
\label{peakiness}

One benefit of the collar-aware loss-function discussed above was the "peaky" output of the model. Figure \ref{fig:peakiness} demonstrates this effect by visualizing samples obtained
from Resnet+BLSTM model outputs centered around randomly chosen annotated boundaries for the English dataset. The output obtained from a model trained using the neighborhood-based  method is spread out over multiple frames requiring finding the exact local maximum in post-processing. In comparison, the change points predicted by the model trained with the collar-based method can be obtained by simply comparing the model outputs to a threshold since the activations tend to be limited to a single frame.

\subsubsection{Tuning the collar size}
\label{tuning}

Figure \ref{fig:tuning} shows the influence of the collar size used during training on the F1 score on English test data. The optimal size for the collar is dependant on the nature of the data, how imbalanced it is and how reliable the annotated boundaries are. Overall, there seems to be flexibility to the choice of collar size as the F1 score does not change a lot across the tested range. Notably, all of the tested collar sizes lead to a better result than the neighborhood based models.

\section{Conclusion}

This work presented a novel supervision method for speaker change detection models using a collar-aware objective function. In our experiments we compared it with a conventional training method that artificially labels a neighborhood of an annotated boundary as positive as well as various state-of-the-art speaker diarization models. We find that our collar-aware training yields improved results both for a purely LSTM-based model and one that uses pretrained embeddings with 8-fold subsampling. 

We analyzed model outputs around randomly chosen boundaries to show that the activations for our method are concentrated to a single frame. This makes our training method well suited for online applications as there is no need for local maxima detection in post-processing.

The exact choice of collar size was determined to not have a great effect on performance with choices from 80ms to 500ms all outperforming the conventional training method.

\section{Acknowledgements}

The authors acknowledge the TalTech supercomputing resources made available for conducting the research reported in this paper.

% References should be produced using the bibtex program from suitable
% BiBTeX files (here: strings, refs, manuals). The IEEEbib.bst bibliography
% style file from IEEE produces unsorted bibliography list.
% -------------------------------------------------------------------------
\bibliographystyle{IEEEbib}
\bibliography{refs}

\begin{thebibliography}{10}

\bibitem{saon2013speaker}
George Saon, Hagen Soltau, David Nahamoo, and Michael Picheny,
\newblock ``Speaker adaptation of neural network acoustic models using
  i-vectors,''
\newblock in {\em ASRU}, 2013.

\bibitem{aronowitz2020context}
Hagai Aronowitz and Weizhong Zhu,
\newblock ``Context and uncertainty modeling for online speaker change
  detection,''
\newblock in {\em ICASSP}, 2020.

\bibitem{india2017lstm}
Miquel~{\`A}ngel India~Massana, Jos{\'e}~Adri{\'a}n Rodr{\'\i}guez~Fonollosa,
  and Francisco~Javier Hernando~Peric{\'a}s,
\newblock ``{LSTM} neural network-based speaker segmentation using acoustic and
  language modelling,''
\newblock in {\em Interspeech}, 2017.

\bibitem{yin2017speaker}
Ruiqing Yin, Herv{\'e} Bredin, and Claude Barras,
\newblock ``Speaker change detection in broadcast {TV} using bidirectional long
  short-term memory networks,''
\newblock in {\em Interspeech}, 2017.

\bibitem{hruz2016convolutional}
Marek Hr{\'u}z and Marie Kune{\v{s}}ov{\'a},
\newblock ``Convolutional neural network in the task of speaker change
  detection,''
\newblock in {\em International Conference on Speech and Computer}, 2016.

\bibitem{hruz2017convolutional}
Marek Hr{\'u}z and Zbyn{\v{e}}k Zaj{\'\i}c,
\newblock ``Convolutional neural network for speaker change detection in
  telephone speaker diarization system,''
\newblock in {\em ICASSP}, 2017.

\bibitem{mateju2019approach}
Lukas Mateju, Petr Cerva, and Jindrich Zd{\'a}nsk{\`y},
\newblock ``An approach to online speaker change point detection using {DNNs}
  and {WFSTs},''
\newblock in {\em Interspeech}, 2019.

\bibitem{rouvier2013open}
Mickael Rouvier, Gr{\'e}gor Dupuy, Paul Gay, Elie Khoury, Teva Merlin, and
  Sylvain Meignier,
\newblock ``An open-source state-of-the-art toolbox for broadcast news
  diarization,''
\newblock in {\em Interspeech}, 2013.

\bibitem{meignier2001hmm}
Sylvain Meignier, Jean-Fran{\c{c}}ois Bonastre, and St{\'e}phane Igounet,
\newblock ``{E-HMM} approach for learning and adapting sound models for speaker
  indexing,''
\newblock in {\em 2001: A Speaker Odyssey-The Speaker Recognition Workshop},
  2001.

\bibitem{malegaonkar2007efficient}
Amit~S Malegaonkar, Aladdin~M Ariyaeeinia, and Perasiriyan Sivakumaran,
\newblock ``Efficient speaker change detection using adapted {Gaussian} mixture
  models,''
\newblock {\em IEEE Transactions on Audio, Speech, and Language Processing},
  vol. 15, no. 6, pp. 1859--1869, 2007.

\bibitem{castaldo2008stream}
Fabio Castaldo, Daniele Colibro, Emanuele Dalmasso, Pietro Laface, and Claudio
  Vair,
\newblock ``Stream-based speaker segmentation using speaker factors and
  eigenvoices,''
\newblock in {\em ICASSP}, 2008.

\bibitem{gupta2015speaker}
Vishwa Gupta,
\newblock ``Speaker change point detection using deep neural nets,''
\newblock in {\em ICASSP}, 2015.

\bibitem{sari2019pre}
Leda Sar{\i}, Samuel Thomas, Mark Hasegawa-Johnson, and Michael Picheny,
\newblock ``Pre-training of speaker embeddings for low-latency speaker change
  detection in broadcast news,''
\newblock in {\em ICASSP}, 2019.

\bibitem{zeghidour2021dive}
Neil Zeghidour, Olivier Teboul, and David Grangier,
\newblock ``{DIVE}: End-to-end speech diarization via iterative speaker
  embedding,''
\newblock {\em arXiv:2105.13802}, 2021.

\bibitem{hannun2017sequence}
Awni Hannun,
\newblock ``Sequence modeling with {CTC},''
\newblock {\em Distill}, vol. 2, no. 11, 2017.

\bibitem{hub4_a}
``{1996 English Broadcast News Speech (HUB4)},''
  https://catalog.ldc.upenn.edu/LDC97S44.

\bibitem{hub4_b}
``{1996 English Broadcast News Transcripts (HUB4)},''
  https://catalog.ldc.upenn.edu/LDC97T22.

\bibitem{Chung18b}
J.~S. Chung, A.~Nagrani, and A.~Zisserman,
\newblock ``{VoxCeleb2}: Deep speaker recognition,''
\newblock in {\em Interspeech}, 2018.

\bibitem{alumae2020taltech}
Tanel Alum{\"a}e,
\newblock ``The {TalTech} system for the {VoxCeleb Speaker Recognition
  Challenge 2020},''
\newblock Tech. {R}ep., 2020.

\bibitem{musan2015}
David Snyder, Guoguo Chen, and Daniel Povey,
\newblock ``{MUSAN}: {A} {M}usic, {S}peech, and {N}oise {C}orpus,'' 2015,
\newblock arXiv:1510.08484v1.

\bibitem{ko2017study}
Tom Ko, Vijayaditya Peddinti, Daniel Povey, Michael~L Seltzer, and Sanjeev
  Khudanpur,
\newblock ``A study on data augmentation of reverberant speech for robust
  speech recognition,''
\newblock in {\em ICASSP}, 2017, pp. 5220--5224.

\bibitem{szoke2019}
I.~{Szöke}, M.~{Skácel}, L.~{Mošner}, J.~{Paliesek}, and J.~{Černocký},
\newblock ``Building and evaluation of a real room impulse response dataset,''
\newblock {\em IEEE Journal of Selected Topics in Signal Processing}, vol. 13,
  no. 4, pp. 863--876, 2019.

\bibitem{landini2022bayesian}
Federico Landini, J{\'a}n Profant, Mireia Diez, and Luk{\'a}{\v{s}} Burget,
\newblock ``Bayesian {HMM} clustering of x-vector sequences ({VBx}) in speaker
  diarization: theory, implementation and analysis on standard tasks,''
\newblock {\em Computer Speech \& Language}, vol. 71, pp. 101254, 2022.

\bibitem{nagrani2017voxceleb}
A~Nagrani, JS~Chung, and AP~Zisserman,
\newblock ``{VoxCeleb}: a large-scale speaker identification dataset,''
\newblock in {\em Interspeech}, 2017.

\bibitem{fan2020cn}
Yue Fan, JW~Kang, LT~Li, KC~Li, HL~Chen, ST~Cheng, PY~Zhang, ZY~Zhou, YQ~Cai,
  and Dong Wang,
\newblock ``{CN-CELEB}: a challenging {Chinese} speaker recognition dataset,''
\newblock in {\em ICASSP}. IEEE, 2020, pp. 7604--7608.

\bibitem{garcia2011analysis}
Daniel Garcia-Romero and Carol~Y Espy-Wilson,
\newblock ``Analysis of i-vector length normalization in speaker recognition
  systems,''
\newblock in {\em Interspeech}, 2011.

\bibitem{bredin2021}
Herv{\'e} {Bredin} and Antoine {Laurent},
\newblock ``End-to-end speaker segmentation for overlap-aware resegmentation,''
\newblock in {\em Interspeech}, 2021.

\bibitem{fujita2019end}
Yusuke Fujita, Naoyuki Kanda, Shota Horiguchi, Kenji Nagamatsu, and Shinji
  Watanabe,
\newblock ``End-to-end neural speaker diarization with permutation-free
  objectives,''
\newblock in {\em Interspeech}, 2019.

\bibitem{ryant2020third}
Neville Ryant, Prachi Singh, Venkat Krishnamohan, Rajat Varma, Kenneth Church,
  Christopher Cieri, Jun Du, Sriram Ganapathy, and Mark Liberman,
\newblock ``The third {DIHARD} diarization challenge,''
\newblock {\em arXiv preprint arXiv:2012.01477}, 2020.

\bibitem{ryant2020third-b}
Neville Ryant, Kenneth Church, Christopher Cieri, Jun Du, Sriram Ganapathy, and
  Mark Liberman,
\newblock ``Third {DIHARD} challenge evaluation plan,''
\newblock {\em arXiv preprint arXiv:2006.05815}, 2020.

\bibitem{coria2021overlap}
Juan~Manuel Coria, Herv{\'e} Bredin, Sahar Ghannay, and Sophie Rosset,
\newblock ``Overlap-aware low-latency online speaker diarization based on
  end-to-end local segmentation,''
\newblock in {\em IEEE Automatic Speech Recognition and Understanding
  Workshop}, 2021.

\end{thebibliography}

\end{document}